# Engineering Predictions in Industrial and Power Flows Using the Retrograde Condensation Curve. Part I-Methodology

Mark S. LABINOV, PhD, PE, CEM

Adjunct-Professor
College of Engineering, Technology and Architecture, University of Hartford,
Bloomfield, CT, USA
E-mail:labinov@hartford.edu

**Abstract**

Industrial and power systems rely on engineering predictions of the flow properties of working fluids. The paper proposes a way of the utilization of the vapor quality values along the retrograde condensation curve in a)generation of the void fraction design guidelines and b)reliable prediction of the saturated liquid specific volumes/densities. The new procedure eliminates the involvement of semi-empirical relationships like rectilinear diameter (Gerais 2010) and other similar models. Retrograde condensation manifests in a pure fluid as a reverse behavior of the vapor quality at a constant specific volume lower then the critical value. For every constant specific volume of that kind there exists only one point where the rate of the quality variation with temperature dx/dT changes its sign from positive to negative. The formula for the locus connecting those points defines the retrograde condensation curve. Unlike all other thermodynamic curves this curve establishes a connection between the thermo-physical saturation properties and the mass property of the process.

## 1. The Retrograde Condensation curve

The existence of the RC-curve was postulated in the series of the research papers by Labinov (1988, 1991,2009) Figure 1 illustrates the concept. For every constant – specific volume process to the left of the critical volume there exists only one unique point at which the rate of quality per temperature changes its sign. When the system is heated, the quality increases until at the particular temperature it starts to decrease; the system crosses the saturation line into the area of superheated fluid. The opposite happens when the system is cooled; it crosses the saturation line into the two-phase zone, while the quality first increases and then starts to decrease after passing the specific point. The locus of these extreme points is defined as an Retrograde Condensation (RC) curve.

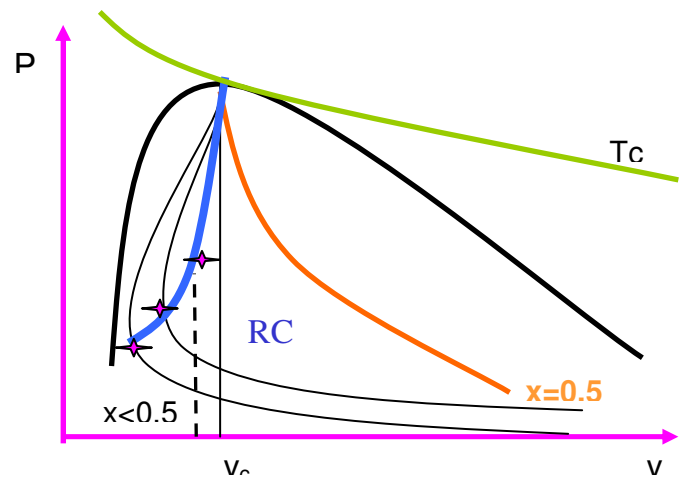

Figure 1

## Thermodynamic equation

The equation for the locus of points where (dx/dT) =0 can be derived from the isochoric process equation (Akopian 1955):

$$v_{t,ph} = v_{vp} x + v_l (1-x)$$

$$\frac{dv_{t,ph}}{dT} = \frac{dv_{vp}}{dT} x + v_{vp} \frac{dx}{dT} + \frac{dv_l}{dT} - x \frac{dv_l}{dT} - v_l \frac{dx}{dT}$$

where under the conditions :

$$dx/dT = 0; dv_{t,ph}/dT = 0$$

we obtain

$$x_{RC} = \frac{dv_l / dT}{[dv_l / dT - dv_{vp}/dT]} \quad (1)$$

In the vicinity of the critical point per the classic theory of the saturation line (e.g., Labinov, 2009) the equation (1) transforms into:

$$\rho_l - \rho_c \approx (T_c - T)^\beta$$
$$\rho_c - \rho_{vp} \approx (T_c - T)^\beta$$
$$x_{RC} \approx [1 + (\frac{\rho_l}{\rho_{vp}})^2]^{-1} \quad (2)$$
$$\lim_{T \to T_c} x_{RC} = 0.5$$

Thus the RC-line asymptotically approaches the x=0.5 at the critical point. Computational experiments conducted with series of hydrocarbons by Labinov, Syssoev and Chaly in 1988 came up with the critical index equation:

$$(1-x)_{RC} = 0.5(1+\eta^\lambda) \quad (3)$$
$$\lambda \approx 0.3$$
$$\eta = 1 - \frac{T}{T_c}$$
$$T/Tc >= 0.7$$

Figure 2 presents a recalculated RC-curve for water (ASME Steam Tables 1993) with the exclusion of the near-critical area.

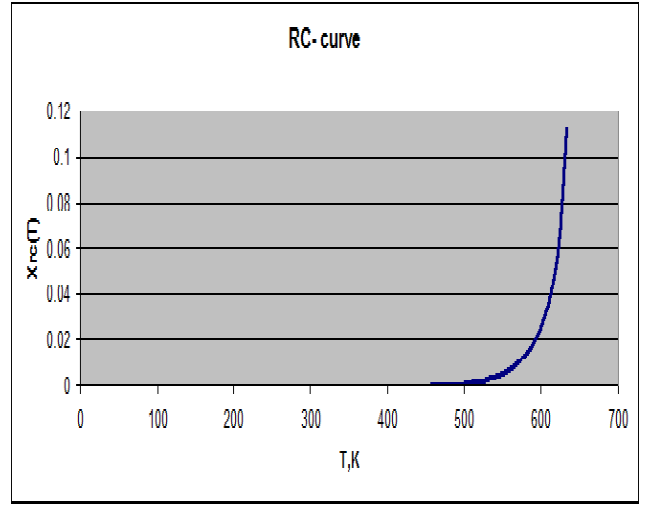

Figure 2

## 2. Void fraction border line

Void fraction, the volumetric fraction of the gas phase in the two-phase flow ε is the primary parameter of the two – phase flow, used extensively in the power systems design and analysis calculations. Numerous correlations connect the void fraction with the vapor quality as well as with the hydrodynamic parameters (slippage, etc.) In 2007 Woldesmayat and Ghajar considered 68 correlations with the purpose to develop a universal correlation. The correlations included primarily quality, density ratios, viscosity ratios, etc.
Among others they listed the practical, highly utilized first- order formula was once successfully postulated by Fauske in 1961:

$$\varepsilon = [1 + \frac{1-x}{x}(\frac{\rho_{vp}}{\rho_l})^{0.5}]^{-1} \quad (4)$$

If we replace the quality x in the Fauske equation (4) with the RC- quality, $x_{Rc}$, we will obtain the void fraction curve on the Figure 3, depending explicitly on temperature – the margin curve for the flow of the pure fluid.

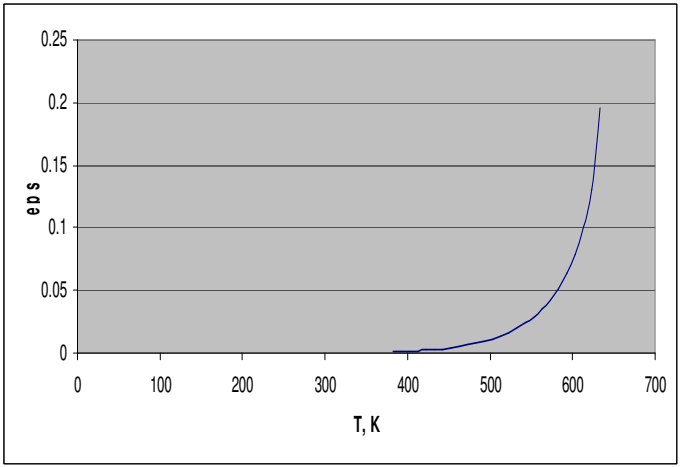

Figure 3

The void fraction line on the Figure.3 thus transfigures the RC- border line between the stable and unstable equilibrium zones defined explicitly on the Figure.4, onto the limitations of the usage of the steady- state- based correlations (heat transfer, friction, etc.) for the two–phase flow.

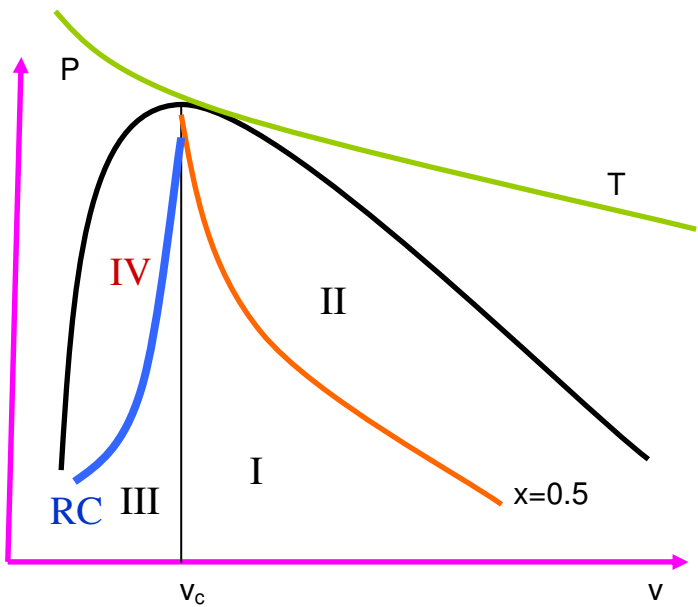

Figure 4

Two-phase constant – volume system is a dynamic system where motion is defined by the change of vapor quality vs. temperature.. Per Le-Chatelier's principle the equilibrium is stable in the regions I, II and III on the Figure 4. In these regions the system's response to the external disturbance (e.g. heat impulse) would be to restore the equilibrium (dx/dT >0; the system absorbs the heat by evaporation). The regions I and II are defined as 'severely damped'; when decaying oscillations occur there, they die up very quickly. Region III is a 'soft damping' region, where decaying oscillations can have low decrement. Per the same principle region IV is the unstable one. System response in this region is to amplify the external influence (dx/dT<0; the system reacts by the retrograde condensation to the heat impulse. The RC-locus is a border line between the regions of stable and unstable equilibrium. When described in terms of the void fraction that means that entering the region IV, especially in the transient cases, can affect the primary design parameters like Net Pumping Suction Head, also protection from the water hammer. The border void fraction curve on the Figure 3 this way becomes a part of an engineering design standard work for the systems with the two-phase flow.

### 3. Liquid density measurements

The Clapeyron – Clausius equation ( Reid, Paruznits and. Sherwood 1977) connects the thermo-physical and phase- transition properties in the two-phase zone:

$$T * DP/DT = \frac{L}{(v_{vp} - v_l)} \quad (5)$$

Differentiation of (5) and combining it with (1) gives the direct equation for the saturated liquid specific volume vs. temperature:

$$dP/DT + T\frac{d^2P}{dT^2} - \frac{dL/dT)}{(v_{vp}-v_l)} = \frac{(L/x_{Rc})*\frac{dv_l}{dT}}{(v_{vp}-v_l)^2} \quad (6)$$

*where*:

$v_{vp} - v_l -$ *taken from* (5)

Equation (6) is the basis for the usage of the mass property (quality) measurements to obtain the thermo-physical property (saturated liquid specific volume or density). The methodology is commensurate to the flow systems' requirements (Wakeham et al. 2007), where instead of the tabulated high-precision data the local limited on–site set of measurements are sought; easier to obtain but still precise enough for the engineering design.
It renders as obsolete the semi-empirical models connecting the saturated vapor and liquid densities, (e.g rectilinear diameter, Gerais 2010).

**The On-site measurement procedure**

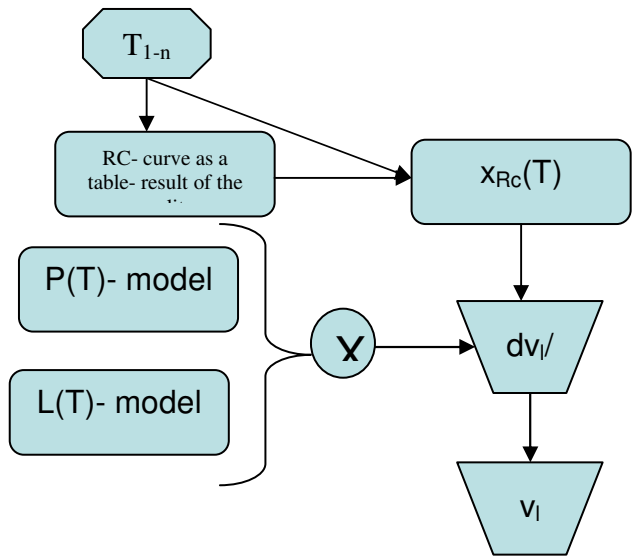

Figure 5

The models for L and P on the Figure 5 are historically well–developed and precise for pure fluids and mixtures (Reid, Prauznitz and Sherwood 1977). Direct vapor quality measurements can be performed either through a flow method (Dorfman et al. 2007 or or by an optical method (Donaldson 1998). The reconstructed RC- quality data ( e.g . using splines like in the book by De Boor 1980) in concert with the liquid specific volumes /densities derived can be utilized further in checking void fraction correlations and the borderline definition described above.

**The RC- Curve as a result of the separate vapor quality measurement .**

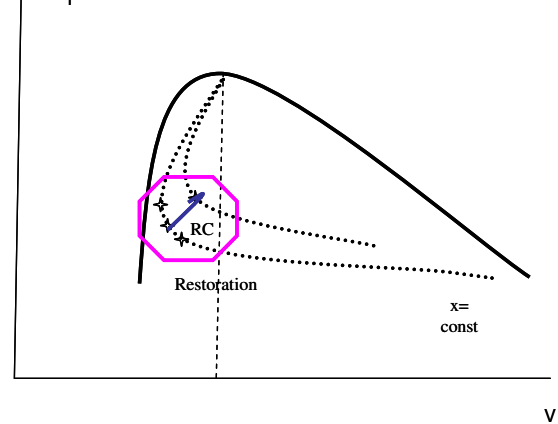

**Figure 6.**

Vapor quality measurements are to be performed as T- x matrices, considering that every x=const corresponds to two T- values to the left of x=0.5. The 'restoration' area is to be treated with splines, (Chebushev polynomials, 1952) or through the detailed testing with reduced temperature step to outline the RC- curve points. Proper result should e the RC- curve provided for the fluid in question with the accompanying average $U_{95}$- uncertainty per ISO – defined methodology (Dieck 2002).

## 4. Closure: measurement uncertainties

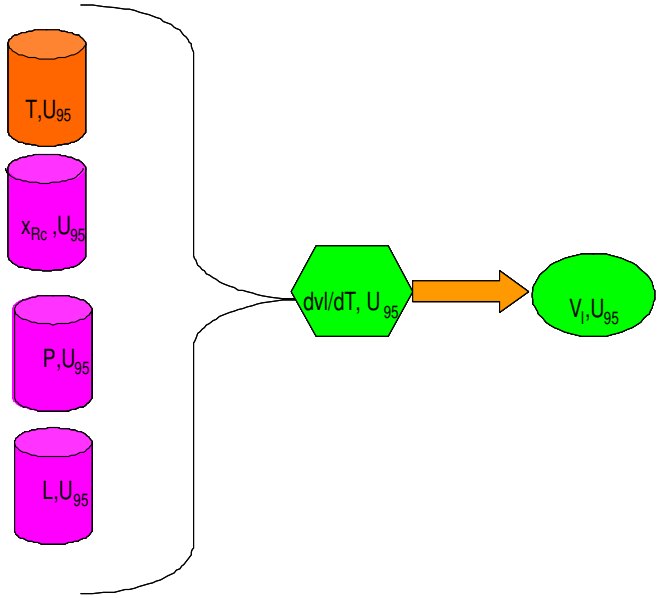

Figure 7

The RC- curve is a newly postulated curve in the two-phase zone and its existence presents new opportunities in the area of flow measurements. Its primary advantage is in the elimination of the semi-empirical models like rectilinear diameter in measurements while retaining the reliability of the classical modeling instead. The schematics on the Figure 7 reveals that there is one permanent in – situ test- related uncertainty component in the proposed measurement process - the uncertainty of the temperature measurements, traditionally of a very high precision. The P, L- models are of high precision also due to historically verified, precise modeling techniques. The lower precision of the $x_{Rc}$- restoration can be compensated by the fact that integration usually leads to higher level of precision ( Guter and Ovcchinski 1970). It is thus conceivable to align the uncertainties in a way that to obtain the $v_l$-predictive relative $U_{95}$ of +-3% , the one industry considers as acceptable (Wakehaam et al. 2007). It is the author's intention to follow this Part I-article with the Part II- an example of the simulated in- situ experiment with the outline of the propagation uncertainty numerics for the known and tabulated group of pure fluids..

## Nomenclature

$T - temperature, (K)$

$G_{t.ph} - mass\ of\ the\ two-phase\ volume, (kg)$

$v_{t.ph} - specific\ two-phase\ volume, (m^3/kg)$

$v_l - specific\ volume\ of\ the\ saturated\ liquid, (m^3/kg)$

$v_{vp} - specific\ volume\ of\ the\ saturated\ vapor, (m^3/kg)$

$v_c - specific\ critical\ volume, (m^3/kg)$

$\rho_l - density\ of\ the\ liquid\ phase, (kg/m^3)$

$\rho_{vp} - density\ of\ the\ saturated\ vapor\ phase, (kg/m^3)$

$\rho_c - critical\ density, (kg/m^3)$

$x = \dfrac{G_{vp}}{G_{t.ph}} - vapor\ quality\ (dimensionless)$

$x_{RC} - quality\ value\ on\ RC-curve\ (dimensionless)$

$G_{vp}, G_l - masses\ of\ the\ vapor\ and\ liquid\ phases, respectively, (kg)$

$P - saturation\ pressure, (Pa)$

$L - heat\ of\ vaporizaiton, (J/kg)$

$\varepsilon - void\ fraction, volumetric\ (dimensionless)$

$U_{95} - here\ absolute\ or\ relative\ 2-sigma\ uncertainty$